# Opportunities and challenges of mRNA technologies in development of Dengue Virus Vaccine

Xiaoyang Liu[1], Daniel Salmon[1*]
[1] Department of International Health, Bloomberg School of Public Health, Johns Hopkins University, Baltimore, MD, United States of America
* Corresponding author
Email:dsalmon1@jhu.edu

**Abstract**
Dengue virus (DENV) is a mosquito-borne virus with a significant human health concern. With 390 million infections annually and 96 million showing clinical symptoms, severe dengue can lead to life-threatening conditions like dengue hemorrhagic fever (DHF) and dengue shock syndrome (DSS). The only FDA-approved vaccine, Dengvaxia, has limitations due to antibody-dependent enhancement (ADE), necessitating careful administration. The recent pre-approval of TAK-003 by WHO in 2024 highlights ongoing efforts to improve vaccine options. This review explores recent advancements in dengue vaccine development, emphasizing potential utility of mRNA-based vaccines. By examining current clinical trial data and innovations, we aim to identify promising strategies to address the limitations of existing vaccines and enhance global dengue prevention efforts.

## 1 Introduction

Dengue virus, a mosquito-borne flavivirus, poses a significant threat to global health, particularly in tropical areas such as the South America, especially the Caribbean including Puerto Rico, the Western Pacific, Eastern Mediterranean, Africa, and the Americas.[1] The World Health Organization (WHO) estimates that there are approximately 390 million dengue infections occur annually, with around 96 million patients manifesting clinical symptoms. Most dengue infections are asymptomatic; however, severe dengue can be lethal. Dengue fever ranges from mild flu-like illness to severe conditions, such as dengue hemorrhagic fever (DHF) and dengue shock syndrome (DSS).[2] DHF is characterized by increased vascular permeability leading to plasma leakage, bleeding, and low platelet count. When not managed promptly, DHF can easily progress to DSS, which is marked by severe hypotension and circulatory collapse neceistating ventilator and pressor support. From 2000 to 2019, the number of confirmed cases reported to WHO has already increased by tenfold. Such a large-scale increase in dengue infection is primarily attributed to more prevalent distribution of disease vectors- chiefly Ades aegypti and Ades albopictus mosquitoes- due to climate change. The impact of climate change on dengue infection is particularly notable in countries previously considered dengue naïve.[3]

Currently, the only FDA-approved dengue vaccine is Dengvaxia (CYD-TDV), developed by Sanofi Pasteur. Approved in 2015, Dengvaxia is a live attenuated tetravalent vaccine designed to protect against all four dengue virus serotypes (DENV-1, -2, -3, and -4).[4] While Dengvaxia represents a significant advance in dengue prevention, it has notable drawbacks. One major concern is the risk of antibody-dependent enhancement (ADE). ADE occurs when non-neutralizing antibodies from a previous dengue infection or vaccination enhance the uptake of the virus of a different serotype into host cells, leading to increased viral replication and more severe symptoms.[5] ADE is concerning in flavivirus-naïve individuals, particularly children under the age of five, as vaccination could potentially predispose them to severe dengue if they contract the virus later. Therefore, Dengvaxia administration is restricted to people who have previously infected with dengue. In this May 2024, another candidate, TAK-003, was pre-approved by WHO.[6] Though TAK-003 will not induce any ADE response, the generated antibody titer against DENV3 was much lower than other strains. Therefore, the need for a more comprehensive, and safe vaccine against dengue still exists. In addition, deployment of dengue vaccine on a global scale still remains limited, with only a small percentage of the world population vaccinated, predominantly in dengue-endemic countries. Such limitation underscores the urgent need for new vaccine strategies that can provide safe and effective protection against all four virus strains without the introduction of ADE. In addition to conventional vaccine technology, mRNA vaccines offer researchers a more versatile platform for rapid vaccine development. Over the past decades, technological innovatios had made mRNA a safe and effective tool in vaccinology, and its successes in treating and prevention infectious diseases, such as rabbies, Zika virus, and influenza virus, have garnered significant attention. mRNA vacccines utilize genetically engineered molecules to eduate the receipients' cells to make a harmless piece of protein that will trigger self-immune system to produce antibody against a specific virus.[36] Since it is non-infectious with high efficacy, mRNA vaccines are safer and can elicit potent immune response faster. In this review, we aim to highlight recent advancement in dengue vaccine development by summarizing recent clinical trial data on dengue vaccine candidates with a particular focus on mRNA DENV vaccine development. By examining the latest research and technological innovations in the field, this review seeks to underscore promising avenues for overcoming the limitations of existing vaccines and improving global dengue prevention efforts.

## 2 Dengue Epidemiology

The earlist documented dengue outbreak dates to 1779, with cases reported in Jakarta, Indonesia, and Egypt.[7] In recent decades, dengue incidence has dramatically increased, putting nearly half the global population at risk. In the last two decades, WHO has seen a significant rise in reported cases and deaths from dengue fever. Reported cases surged from 505,430 in 2000 to over 2.4 million in 2010 and exceeded 5.2 million in 2019. Similarly, reported deaths increased from 960 in 2000 to 4,032 in 2015.[8] It is estimated that annually, around 390 million infections of Dengue virus (DENV) may occur, with 67–136 million cases exhibiting clinical symptoms and leading to severe illness or death, especially in tropical and subtropical regions. In the United States, dengue fever incidence has historically been low, with recent spikes occurring between 2013 and 2016 (0.17-0.31 cases per 100,000) and peaking in 2019 (0.35 cases per 100,000). 94% of cases between 2010 and 2021 were travel related. However, Puerto Rico has seen a higher incidence, with an average of 200 cases per 100,000 between 1980 and 2015, and almost all cases were locally acquired.[9] The substantial increase in dengue infections is primarily attributed to the expanded distribution of its vectors, particularly Aedes aegypti and Aedes albopictus mosquitoes, driven by climate change. As temperature rises within the survival range for the vector mosquitos, their reproduction will be enhanced, the annual reproductivity period will be elongated, and the virus incubation period will also be shortened. Additionally, increased rainfall, flooding, and humidity also contribute to the breeding of mosquitos.[10,11] Various social and environmental factors, such as population density increase, population mobility, and inadequate water storage practices, are closely associated with dengue transmission, particularly in rural areas.[12] According to a longitudinal study on the dynamics of mosquitos and dengue incidences from 2007 to 2008 in Puerto Rico, water storage containers and discarded tires play vital roles in mosquito incubation, as most pupae were found in human-managed water constrainers, storage vessels, plant pots, and leaking water meters.[13]

## 3 Clincially Approved vaccines

### 3.1 Dengvaxia

Invented by Sanofi Pasteur, Dengvaxia (CYD-TDV) is a live-attenuated, tetravalent vaccine administered as a three-dose regimen. It was approved by the U.S. Food and Drug Administration (FDA) in 2019. Incorporating structural pre-membrane (prM) and envelope (E) genes of the four DENV strains with non-structural genes of the yellow fever 17D vaccine strain, Dengvaxia aims to provide tetravalent immunity to all strains of DENV by targeting the prM and E proteins.[14] The general vaccine efficacy (VE) reached 91% (95% CI, 58-99%) for populations greater than 9 years of age who are sero-positive to dengue (Table 1). For patients less than 9 years of age and who are naïve to dengue, VE was approximately 45% (95% CI, -54-88%), indicating very poor performance in pediatric population.[15] Therefore, Dengvaxia works as a "fill-in" vaccine with patients' prior exposure served as the initial prime. The VE against serious dengue diseases can reach as high as 93.2% for recipients with prior dengue infection.[16] In a phase-II clinical trial (NCT00880893), the safety of dengvaxia was evaluated in subjects aged 2 to 45 years in Singapore. Throughout the whole study, there were only three (0.3%) recorded serious adverse events (SAEs) in the vaccination group- acute leukemia of ambiguous lineage, tuberculosis lymphadenitis, and tension headache- and three cases of adverse events (AEs)- fever, rash, and cervical spondylosis.[17] On the other hand, Dengvaxia may present as a safety concern for dengue-naïve subjects and is not recommended for administration in pediatric populations under nine who are less likely to have a prior dengue infection.

### 3.2 TAK-003

TAK-003 (Takeda) is a live-attenuated, tetravalent vaccine, comprising of four DENV strains with the attenuated DENV serotype 2 strain (DENV-2) as the vaccine backbone and three other recombinant strains, swapping the prM and E genes of DENV-2 with DENV-1, DENV-3, or DENV-4 (Table 1).[18] Previous phase I and II clinical trials for TAK-003 vaccine had addressed its capability of eliciting tetravalent neutralizing antibody responses and polyfunctional T-cell responses. In a phase III trial (NCT02747927), two doses of TAK-003 administered to children 4 to 16 years of age showed a general VE of 66.2% (95% CI, 44.9-77.5) for dengue seronegative recipients and 76.1% (95% CI, 68.5-81.9) for seropositive subjects.[19] Cumulatively, VE was 90.4% (95% CI, 82.6-94.7) and 85.9% (95% CI, 31.9-97.1) against hospitalizations related to dengue and DHF. After three-year administration of TAK-003, the cumulative VE dropped to 54.3% (95%CI, 41.9-64.1) for virologically confirmed dengue (VCD) and 77.1% (95%CI, 58.6-87.3) against hospitalized VCD in the baseline seronegative cohort. TAK-003 maintained its efficacy in the baseline seropositives cohorts, with 65% (95%CI, 58.9-70.1) VE against VCD and 86% (95%CI, 78.4-91.0) against hospitalized VCD. In another phase III trial evaluating safety of TAK-003 (NCT03771963), among 168 participants, there were only five participants experienced SAEs with two subjects reporting moderate hepatic failure and severe osteoarthritis and three reporting bradycardia, inguinal hernia, and sepsis.[20] However, there was no efficacy observed against DENV-3 in the baseline seronegative volunteers, and VE against DENV-2 wanned overtime, raising possibilities of ADE from DENV-2 and DENV-3 infections years after immunizations.

**Table 1.**
**Current licensed or trialed Dengue vaccines**

| Name | Valency | Formulation | Evaluation | Manufacturer | Efficacy | Adverse events | Dose Schedule | Year |
|---|---|---|---|---|---|---|---|---|
| Dengvaxia | Tetravalence | Chimeric combination of | Licensed | Sanofi Pasteur | The general VE[e] against all four | ADE response occurred in dengue | 3 | 2015 |
| | | YFV/DENV 1-4 | | | serotypes was 65.6% | naïve individuals was the major safety concern | | |
| TAK-003 | Tetravalence | Chimeric viruses with DENV-2 PDK35 as the backbone | Pre-licensed on May 2024 by WHO | Takeda | The cumulative VE against DENV1-4 was 66.2% | The most common adverse events were injection site pain and headache | 2 | 2006 |
| TV003/TV005 | Tetravalence | Genetic attenuated viruses | Phase-III clinical trial | NIAID[a], Butantan, and Merck | Seroconversion rate[f] for TV003 was 74% and 97% for TV005 | Headache, rash, fatigue, and myalgia were the most common observed adverse events | 1 | 2003 |
| TDEN F17/F19 | Tetravalence | Virus combination attenuated by PDK cells | Phase II clinical trial | WRAIR[b] and GSK[c] | Seroconversion rate against DENV1-3 was 100% and 83.3% against DENV4 | Arthralgia, fatigue, muscle aches, and pain behind eyes were observed in recipients | 2 | 2017 |
| DPIV | Tetravalence | Purified inactivated DENV1-4 with aluminum adjuvants | Phase I clinical trial | WRAIR and GSK | Tetravalent neutralizing antibodies were induced | There were few cases of moderate adverse events recorded during the trial | 2 | 2012 |
| TVDV | Tetravalence | DNA vaccine encoding prM and E proteins of DENV1-4 and adjuvanted with VAXFECTIN | Phase I clinical trial | WRAIR and U.S. NMRC[d] | Anti-DENV IFN-γ T cells response was stimulated | No severe adverse events were observed | 3 | 2018 |
| V180 | Tetravalence | Recombinant prM and E proteins of DENV1-4 combined with | Phase I clinical trial | Merck & Co. | Seroconversion rate against all four serotypes was 85.7% | Injection site pain was the most common adverse | 3 | 2018 |

| | | | |
|---|---|---|---|
| | multiple adjuvants | event througho ut the trial | |

[a] National Institute of Allergy and Infectious Diseases; [b] Walter Reed Army Institute of Research; [c] GlaxoSmithKline; [d] U.S. Naval Medical Research Center;

[e] VE refers to vaccine efficacy, which is measured by comparing the number of disease cases in the vaccinated group to that of the placebo group;

[f] Seroconversion rate is the percentage of individuals who develop detectable specific antibodies to a pathogen in their blood post vaccination or infection. This rate is a common indicator of vaccine effectiveness in immunological research.

## 4 Vaccines under evaluations in clinical trials

### 4.1 TV003/TV005

The National Institute of Allergy and Infectious Disease (NIAID) has led efforts to develop a live-attenuated tetravalent dengue vaccine over the past 15 years, aiming to create a vaccine that could confer protection against all four strains of dengue virus and address the risks of antibody dependent enhancement. For TV003, with the introduction of a thirty (Δ30) and thirty-one nucleotide (Δ31) deletion at the 3' UTR of the four dengue serotypes, four efficacious monovalent candidates- rDEN1Δ30, rDEN2/4Δ30, rDEN3Δ30/3Δ31- were combined into tetravalent formulations, and rDENV2/4Δ30 was found to be least effective in the tetravalent combinations.[21] To achieve a more balanced infectivity, in the TV005 formulation, the doses of rDENV2/4Δ30 were increased 10 fold.[22] From previous phase I trials (NCT01072786 and NCT01436422), comparing to TV003, TV005 demonstrated a higher and more stable frequency of seroconversion and stronger antibody response (Table 1). Both TV003 and TV005 demonstrated sterilizing immunity against DENV infection for at least 12 months when a second dose was administered 6 months after the first dose. Dengue vaccine related rash was the only observed moderate adverse event for both vaccine candidates. Among reciepients who were vaccinated with TV003, 66% (27/41) experienced this mild rash. TV005 caused rash in 26% (37/144) vaccinees, but seven vaccinees experienced fever and seven recipeints experienced arthralgias.[23] After the second dose vaccination, there was an approximate two-fold increase in the mean antibody titer to all four serotypes, comparing to first dose. This further demonstrated that sterilizing immunity was introduced by TV003/TV005. TV003/TV005 *are currently under Phase* III clinical trial evaluation in dengue-endemic regions (NCT02406729), led by the Brazilian institute Butantan.

### 4.2 TDEN F17/F19

TDEN F17 vaccine is a tetravalent, live-attenuated vaccine that targets all four dengue virus serotypes. The vaccine was dedrived from a natural viral isolate and attenuated through serial passages in primary dog kidney (PDK) cells. In order to induce stronger neutralizing antibody reponses, the F17pre formulation was devised with a more attenuated DENV1 strain by higher PDK cell passage and a less attenuated DENV4 strain by lower PDK cell passage.[24] Clinical trial (NCT00350337) demonstrated that TDEN F17/F17Pre/F19 induced robust humoral responses with tetravalent response rates of 60%, 71.4%, and 66.7% after 2-dose administration, and a phase-II trial conducted in Puerto Rico (NCT00468858) achieved 100% seroconversion to tetravalent immunity in primed subjects.[25, 26] In a pilot study on TDEN F17's VE in children, a 52.6% seroconversion was achievd, suggesting the vaccine might be safe and effective in children. With a two-dose regimen, TDEN F17 is deemed safe across a broad age range- from 12 months to 50 years of age. However, viremia was noted in a significant proportion of subjects after the first dose-31% for formulations F17pre- though absent after the second dose.[27] A five-year-phase I/II studies conducted in Thailand children aged 6 to 7 assessing long-term immunogenicity and safety (NCT00384670) with a one-year-follow-up evaluating the effect of a third dose booster administered one year after the second dose (NCT01843621) demonstrated promising results. Seroconversion among all participants were 100% (95%CI, 54.1-100) to DENV-1, -2, and -3, and 83.3% (95%CI, 35.9-99.6) to DENV-4. There was neither mortality case or SAEs observed in both trials, indicating the safety use of TDEN F17.[28]

### 4.3 DPIV

Invented at Walter Reed Army Institute of Research (WRAIR), the tetravalent dengue purified inactivated vaccine (DPIV) is administered with two-dose schedule 28 days apart. The DENV-2 S16803 strain was chosen as the initial vaccine prototype, which was successfully propagated in Vero cells and tested safe and immunogenic in mice and rhesus monkeys with a 100% seroconversion after the second dose administration.[29] Adjuvanted with $AS01_E$(3-O-desacylcinomnophsphoryl lipid A) and $AS03_B$(oil-in-water) by GlaxoSmithKline, the formalin-inactivated viruses from DENV-1 Westpac 74, DENV-2 S16803, DENV-3 CH53489, and DENV-4 TVP360 were incorporated into the tetravalent formulation of DPIV. In a phase-I trial conducted in Puerto Rico (NCT01702857), DPIV adjuvanted by AS01E/AS03B elicited neutralizing antibody responses against all four DENV serotypes in flavivirus-naïve adults, but the response wanned in 6 months after the second dose.[30] In addition, the inactivated nature of DPIV poses a potential limit on immune responses to non-neutralizing epitopes on target envelope and capsid proteins. Viral challenge studies in rhesus macaques revealed the vaccine failed to control DENV infection and inadvertently led to antibody-dependent enhancement of DENV infection with increased levels of viremia, AST, IL-10, and IL-18 in challenged animals.[31]

### 4.4 TVDV

The tetravalent DNA vaccine against dengue (TBDV) was developed by the U.S. Army Medical Research and Material Command. The vaccine relies on the VAXFECTIN-adjuvated VR1012 plasmid encoding prM and E proteins and is given as a three-dose regimen. Each component in TBDV was derived from prM and E proteins from different serotypes and TBDV combined equal amounts the monovalent plasmids, VR1012. DENV1 component was derived from West Pacific 74 strain, DENV2 section was generated from modified original DENV2 strain, and DENV3 and DENV4 components were extracted from low passage DENV3 and -4 Phillipine strains.[32] When evaluated in New Zealand white rabbits, TBDV induced seroconversions against all four serotypes, and a phase-I clinical trial (NCT01502358) with 40 flavivirus-naïve recipients indicated TVDV was safe and could effectively elicit IFN-γ anti-DENV T cell responses. [32,33]Throughout the trial, no recipients experienced dengue vaccine related rash or SAEs. The most common AEs occurred among participants were fatigue (17/40), headache (18/40), and myalgias (19/40). Though there were no neutralizing antibody responses detected, T cell responses- the accumulation of IFNγ- were induced with a average response rate around 66.3% amongst three subjects group-falvivirus-naïve volunteers received low-dose TVDV alone, low-dose TVDV adjuvanted with Vaxfectin, or high-dose TVDV adjuvanted with Vaxfection.[32]

### 4.5 V180

Developed by Merck & Co., V180 is a recombinant, tetravalent vaccine given as a three-dose regimen that targets DENV envelope and prM glycoproteins. In a phase-I, randomized, placebo-controlled, double-blinded study (NCT01477580), neither unadjuvanted V180 formulations nor aluminum-adjuvanted formulation elicited a robust immune response; however, six V180 formulations combined with ISCOMATRIX adjuvant were capable of eliciting robust immunogenicity (GMT > 150) with a mean seroconversion rate for all four DENV serotypes greater than 85.7%. Memory B cells responses for all dengue serotypes were also detected in participants who received high dose of V180-ISCOMATRIX. Unfortuantely, V180-ISCOMATRIX was noted to cause more AEs, such as injection site pain and swelling.[34] Another phase-I randomized clinical trial (NCT02450838) was conducted in adults who were previously vaccinated with a live-attenuated tetravalent dengue vaccine (LATV) and evaluated imunogencity and safety of V180 formulations, either plain or adjuvanted with Alhydrogen, as a booster. V180 formulations were found to be well tolerated in reciepients and were able to enhance serum neutralization titers; nevertheless, the elicited immunogenicity did not achieve the criteria set for a positive booster immune response (GMT > 150).[35]

## 5 mRNA DENV vaccines

In addition to traditional tetravalent live-attenuated DENV vaccines, the unprecedented success of COVID-19 mRNA vaccines indicates a great potential for DENV mRNA vaccines. Unlike viral vectored vaccines, mRNA vaccines do not contain nucleic acids that can integrate into the host genome, thereby avoiding the risk of potential mutagenesis and oncogenesis. Moreover, mRNA vaccines are cost-effective, can be quickly manufactured and induce cellular and humoral immunity rapidly.[36]

In 2019, a pre-clinical study was conducted by Claude Roth et al in human HLA Class-1 (HLA-A0201, -A2402, and B3501) transgenic mice to demonstrate the efficacy of a modified mRNA vaccine against DENV-1 strain KDH0026A (Figure 1A).[37] This vaccine encoded the immunodominant non-structural epitopes, DENV1-NS, which are enriched in the NS3, NS4B, and NS5 domains to enhance CD4+ and CD8 + T cells response.[38,39] A lipid nanoparticle (LNP) was engineered to encapsulate the modified mRNA vaccine for delivery. After immunizing HLA-1 transgenic mice, strong CD8 T-cell responses were induced, with 26% of CD8 T-cells producing IFN-γ and TNF-α against HLA-B*3501 peptides p49.* Neutralizing antibody was not induced by the vaccine intentionally so as to avoid the possibility of ADE response introduction.[37] However, the study neither investigated immunogenicity of mRNA vaccines against other DENV serotypes nor demonstraed efficacy of the vaccines in transgenic mice challenge model.

In 2020, another modified mRNA-LNP vaccine against DENV-2 was tested in mice (by Mengling Zhang et al). This mRNA vaccine targeted two structural proteins, prME and E80, and one non-structural protein (NS1) against DENV2 strain 16681 (Figure 1B). Volunteers received either prME-mRNA candidate, E80-mRNA candidate, or NS1-mRNA candidate randomly. After vaccination, immunocompetent BALB/c mice showed high levels of DENV2-specific neutralizing antibodies (DENV2 virion-specific IgG), strong T cell responses, and sterilizing immunity against DENV2. E80-mRNA elicited E80 protein-binding antibodies and neutralizing antibodies with an average $PRNT_{50}$ titer of 13,000. NS1-mRNA induced NS1 portein specific antibody at a $PRNT_{50}$ level of 12,000. When NS1-mRNA vaccine administered alone, though recipient mice demonstrated a statistically significant decreased viral load in spleen than the control group after challenge, NS1-mRNA vaccine did not induce neutralizing antibodies. Thus, NS1-mRNA with limited protection capacity can not be utilized as the major compoenent of the mRNA vaccine.[40] Although E80-mRNA showed promise, it also induced high levels of heterologous antibody-dependent enhancement (ADE) and cross-reactive immune responses, limiting its utility.

In the following year, another similar mRNA-LNP vaccine against serotype 1, the DENV1 prM/E mRNA-LNP vaccine, was developed (by Clayton J. Wollner et al) (Figure 1C).[41] Researchers utilized sequences encoding prM and E proteins from DENV1 strain 16007, as the main component of the mRNA constructs. Additionally, this construct contained a T7 promoter sequence for mRNA transcription and 5' and 3' UTRs, which had been already utilized in Zika virus mRNA vaccine.[42] With a two-dose regimen, both humoral and cellular immunity were induced in mice. Using FRNT as the quantification means, the average antibody titers of 120,000 and neutralizing antibody titers of 420 were elicited. Additionally, this mRNA vaccine protected immunocompromised AG129 mice, which lack the type I interferon $\alpha/\beta$ receptor and the type II interferon $\gamma$ receptor, from a lethal DENV challeng3.[43] After vaccination, there was no morbidity or mortality signs shown among these mice, demonstrating a successful protection. This indicated the vaccine's potential in protecting immunocompromised. Although the prM/E mRNA-LNP vaccine elicited CD4+ and CD8+ T cell responses against DENV1, it did not induce cross-reactive T cell responses against other DENV serotypes.

Previous DENV mRNA vaccines only targeted one serotype; fortunately, in 2022, a multi-target mRNA-LNP vaccine formuation was designed and tested in mice (by Lihong He. et al) (Figure 1D).[44] Coated with LNP, these multi-valent mRNA vaccines encode antigens containing NS1 and envelope domain III (E-DIII) domain and which in theory would be capable of blocking all four types of DENV infection. The DENV-a candidate was a combination of serotype 1 and 2, and the DENV-b candidate was a chimera of serotype 3 and 4. The efficacy of DENV-a, DENV-b, and DENV-ab was tested separately in mice. DENV-a elicited E-DIII specific IgG against DENV1 with an average antibody titer of 15,264 with a 20 μg dose. DENV-b immunized mice generated an E-DIII mean titer of 4608 in the 20 μg group. For NS-1, both DENV-a and DENV-b induced slightly higher titers than those of E-DIII, with mean titers of 61,440 and 32,768.[44] After immunization, DENV-ab elicited highest neutralizing antibodies against all four DENV strains, comparing to DENV-a and DENV-b. In addition to neutralizing antibodies, all vaccine candidates activated T lymphocytes and secreted IFN- $\gamma$, with DENV-ab generated the strongest T cells response. Additionally, ADE assays had been performed to evaluate whether NS1 component would elicit ADE response. Results demonstrated that DENV-ab was safe and did not introduce significant ADE against all four DENV strains in mice, with less than 5% of cells having ADE responses.[44] While current mRNA vaccines candidates seemed promising by effectively inducing both cellular and humoral responses in mice, further studies in larger mammals, such as rabbits and non-human primates, would be necessary before these vaccines could be advanced into clinical trials. based on current research on the multi-target mRNA -LNP vaccines, future efforts should be devoted in developing tetravalent mRNA vaccines that can elicit both cellular and humoral responses against all four virus strains without inducing severe health effects, such as ADE response.

Based on current research on the multi-target mRNA -LNP vaccines, future efforts should be devoted in developing tetravalent mRNA vaccines that can elicit both cellular and humoral responses against all four virus strains without inducing severe health effects, such as ADE response. In addition to improving current LNPs formulation, lumazine synthase (LuS), can also be considered to be incorporated into future development of DENV mRNA vaccines.[45] LuS oligomers, displaying as an efficient platform for antigen presentation, have already been successfully employed in other vaccine studies against infectious diseases, such as HIV, influenza, and rotavirus. When utilized as a scaffold, LuS was able to display spike glycoprotein from SARS-CoV-2 with decent yield and antigenicity in mice.[46] Multimerization nature of LuS contributes to assembling antigens in a highly ordered fashion and promoting the activation of B cells receptors; thereby, potent and long-lasting immune responses in the germinal centers will be generated.[47] When utilized as a protein cage, LuS carrying ovalbumin peptides OT-1 and OT-2 efficiently delivered and successfully stimulated Dentric Cells (DCs) to produce OT-1 specific CD8+ T cells and OT-2 specific CD4+ T cells in mice.[48]

**Figure 1. Schematic illustrations of mRNA DENV vaccine designs**

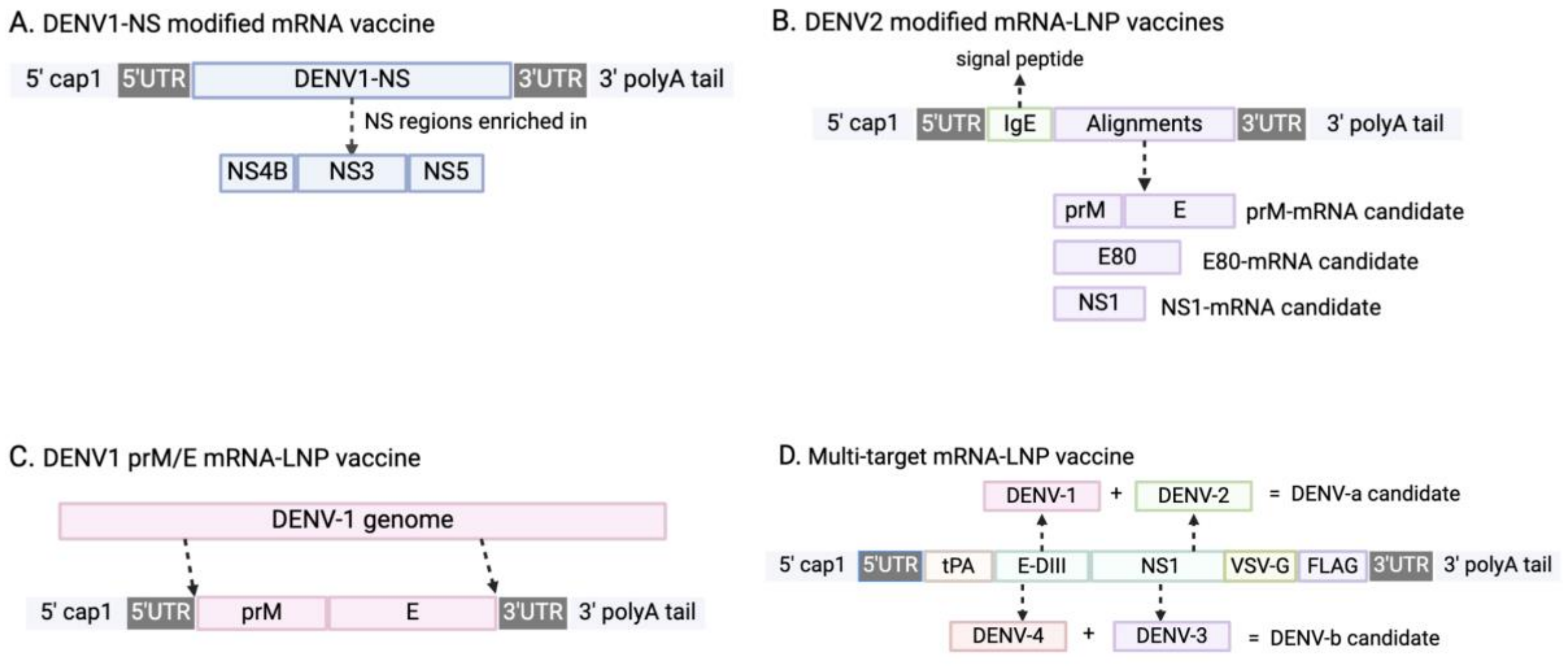

[A.]The major component of DENV1-NS modified mRNA vaccine is non-structural (NS) protein from DENV1 strain, and is enriched in NS4B, NS3, and NS5. [B.] DENV2 modified mRNA-LNP vaccines incorporate a sequence of human IgE as a singal peptide to prompt alignments translocation. [C.]The major component of this vaccine are sequences of pre-membrane(prM) and envolpoe (E) proteins extracted from DENV-1 genome. [D.] The design incorporated tissue-type plasminogen activator (tPA) as the signal peptide, transmembrane and cytoplasmic domain of vesicular stomatitis virus G (VSV-G), and FLAG tag to aid the identification of target protein expression.

## 6 Conclusion

Currently, Dengvaxia is the only DENV vaccine approved by WHO and has been administered worldwide. Fortunately, in May 2024, WHO prequalified another candidate, TAK-003, which expanded the global access to dengue vaccines. Among all existing DENV vaccines that have been advanced to clinical trials, TV003/TV005 were found to elicit the strongest humoral response among flavivirus-naïve recipients, and TV005 induced less adverse events such as vaccine related rashes, comparing to TV003. Based on current data, TV003/TV005 provide at least 1 to 2 years protection with a one-dose regimen, while Dengvaxia and TAK-003 can provide at least 6 years' and 4.5 years' protection after first dose.[49] Among all vaccine candidates, there were only less than 1% of participants experienced SAEs after administration; rather, vaccine related rash was the most common observed side effect among all vaccine candidates that have been mentioned above. ADE response is another significant safety concern for all DENV vaccines. Among all existing DENV vaccines, Dengvaxia is the only one associated with an increased risk of severe dengue in seronegative individuals at the time of vaccination, which is highly likely due to ADE. To mitigate this risk, Dengvaxia is restricted only to individuals who have had a previous dengue infection and serology test will be performed before vaccination.[50] For other candidates, there was a lower risk for ADE based on data from current trials.

mRNA DENV vaccine candidates, on the other hand, hold significant promise due to their rapid development, scalability, precision, and strong immune response. Unlike current live-attenuated dengue vaccines, mRNA DENV vaccines can be developed and manufactured at scale rapidly, as they do not require the growth of pathogens or cells. Furthermore, mRNA DENV vaccines can be designed to encode antigens specific to all four virus strains to induced balanced antibody responses. Based on current preclinical trials, most mRNA vaccine candidates were caplable of inducing neutralizing antibodies and activate T lymphocytes response safely. Multi-target mRNA-LNP vaccines, which target four serotypes and induce sufficient neutralizing antibodies and T cells response, seem to be the most promising candidate so far. Further study on this mRNA vaccines, especially the DENV-ab candidate, should be conducted in Non-Human Primates, such as Rhesus macaque, Cynomolgus macaques, or Common marmosets, to evaluate its safety and immunogenicity.[44] Once the candidate showed safety utilization in NHPs and was able to elicit strong immune responses, it could be advanced to clinical trials and eventually be brought to the market.

Future effort in DENV vaccines development should continue to focus on enhancing their efficacy and safety. One primary area is broadening the immune response, which can be achieved by incorporating adjuvants to current live-attenuated vaccine candidates; for example, aluminum adjuvants, such as MF59 and AS01, can be introduced to current vaccine formulation to enhance Th2 immune responses.[51] Specialized vaccine design is another critical focus area. Self-assembling nanoparticles, which mimic the structure of viruses, enhance antigen presentation, and improve immune response, are a potential candidate. Nanoparticles can be engineered to display multiple dengue antigens that cover all four serotypes, promoting a more robust and comprehensive immune response.[52] In addition, virus-like particles (VLPs) and multiepitope designs can also be employed to present viral antigens in their native conformation, inducing stronger immune responses without the risk of viral replication in recipients.[53,54] For mRNA DENV vaccine candidates, future efforts can be commited to further improvement of the LNP formulation, in addition to neucloetide modification, and developing novel antigen presentation platform, such as LS, to induce more effective and efficient immune response. Furthermore, future efforts should also be devoted in refining dose regimen for different demographic groups in those undergoing Phase II and III clinical trials, as current vaccine candidates are all restricted to children under the age of four and elderly above the age of sixty. It is pivotal to broaden the protection scope to achieve the goal of herd immunity in dengue-endemic areas.

[1] Bhatt, S., Gething, P., Brady, O. *et al.* (2013). The global distribution and burden of dengue. *Nature* 496, 504–507 https://doi.org/10.1038/nature12060

[2] Pintado Silva, J., & Fernandez-Sesma, A. (2023). Challenges on the development of a dengue vaccine: a comprehensive review of the state of the art. *The Journal of general virology*, *104*(3), 001831. https://doi.org/10.1099/jgv.0.001831

[3] World Health Organization (WHO). (2024, April 23). *Dengue and severe dengue*. https://www.who.int/news-room/fact-sheets/detail/dengue-and-severe-dengue

[4] Administration FaD DENGVAXIA. 2020. www.fda.gov/vaccines-blood-biologics/dengvaxia

[5] Teo, A., Tan, H. D., Loy, T., Chia, P. Y., & Chua, C. L. L. (2023). Understanding antibody-dependent enhancement in dengue: Are afucosylated IgG1s a concern?. *PLoS pathogens*, *19*(3), e1011223. https://doi.org/10.1371/journal.ppat.1011223

[6] Takeda Takeda's dengue vaccine candidate provides continued protection against dengue fever through 4.5 years in pivotal clinical trial. 2022. www.takeda.com/newsroom/newsreleases/2022/takedas-dengue-vaccine-candidate-provides-continued-protection-against-dengue-fever-through-4.5-years-in-pivotal-clinical-trial/

[7] Wu, W., Bai, Z., Zhou, H., Tu, Z., Fang, M., Tang, B., Liu, J., Liu, L., Liu, J., & Chen, W. (2011). Molecular epidemiology of dengue viruses in southern China from 1978 to 2006. *Virology journal*, *8*, 322. https://doi.org/10.1186/1743-422X-8-322

[8] World Health Organization (WHO). (2023, December 21). *Dengue: Global situation*. https://www.who.int/emergencies/disease-outbreak-news/item/2023-DON498

[9] Chen, L. H., Marti, C., Diaz Perez, C., Jackson, B. M., Simon, A. M., & Lu, M. (2023). Epidemiology and burden of dengue fever in the United States: a systematic review. *Journal of travel medicine*, *30*(7), taad127. https://doi.org/10.1093/jtm/taad127

[10] Hawley W. A. (1988). The biology of Aedes albopictus. *Journal of the American Mosquito Control Association. Supplement*, *1*, 1–39.

[11] Higa Y. (2011). Dengue Vectors and their Spatial Distribution. *Tropical medicine and health*, *39*(4 Suppl), 17–27. https://doi.org/10.2149/tmh.2011-S04

[12] Schmidt, W. P., Suzuki, M., Thiem, V. D., White, R. G., Tsuzuki, A., Yoshida, L. M., Yanai, H., Haque, U., Tho, leH., Anh, D. D., & Ariyoshi, K. (2011). Population density, water supply, and the risk of dengue fever in Vietnam: cohort study and spatial analysis. *PLoS medicine*, *8*(8), e1001082. https://doi.org/10.1371/journal.pmed.1001082

[13] Barrera R, Amador M, MacKay AJ (2011) Population Dynamics of *Aedes aegypti* and Dengue as Influenced by Weather and Human Behavior in San Juan, Puerto Rico. PLOS Neglected Tropical Diseases 5(12): e1378. https://doi.org/10.1371/journal.pntd.0001378

[14] Guy, B., Briand, O., Lang, J., Saville, M., & Jackson, N. (2015). Development of the Sanofi Pasteur tetravalent dengue vaccine: One more step forward. *Vaccine*, *33*(50), 7100–7111. https://doi.org/10.1016/j.vaccine.2015.09.108

[15] Paz-Bailey, G., Adams, L., Wong, J. M., Poehling, K. A., Chen, W. H., McNally, V., Atmar, R. L., & Waterman, S. H. (2021). Dengue Vaccine: Recommendations of the Advisory Committee on Immunization Practices, United States, 2021. *MMWR. Recommendations and reports : Morbidity and mortality weekly report. Recommendations and reports*, *70*(6), 1–16. https://doi.org/10.15585/mmwr.rr7006a1

[16] Hadinegoro, S. R., Arredondo-García, J. L., Capeding, M. R., Deseda, C., Chotpitayasunondh, T., Dietze, R., Muhammad Ismail, H. I., Reynales, H., Limkittikul, K., Rivera-Medina, D. M., Tran, H. N., Bouckenooghe, A., Chansinghakul, D., Cortés, M., Fanouillere, K., Forrat, R., Frago, C., Gailhardou, S., Jackson, N., Noriega, F., … CYD-TDV Dengue Vaccine Working Group (2015). Efficacy and Long-Term Safety of a Dengue Vaccine in Regions of Endemic Disease. *The New England journal of medicine*, *373*(13), 1195–1206. https://doi.org/10.1056/NEJMoa1506223

[17] Capeding, R. Z., Luna, I. A., Bomasang, E., Lupisan, S., Lang, J., Forrat, R., Wartel, A., & Crevat, D. (2011). Live-attenuated, tetravalent dengue vaccine in children, adolescents and adults in a dengue endemic country: randomized controlled phase I trial in the Philippines. *Vaccine*, *29*(22), 3863–3872. https://doi.org/10.1016/j.vaccine.2011.03.057

[18] Patel, S. S., Winkle, P., Faccin, A., Nordio, F., LeFevre, I., & Tsoukas, C. G. (2023). An open-label, Phase 3 trial of TAK-003, a live attenuated dengue tetravalent vaccine, in healthy US adults: immunogenicity and safety when administered during the second half of a 24-month shelf-life. *Human vaccines & immunotherapeutics*, *19*(2), 2254964. https://doi.org/10.1080/21645515.2023.2254964

[19] Rivera, L., Biswal, S., Sáez-Llorens, X., Reynales, H., López-Medina, E., Borja-Tabora, C., Bravo, L., Sirivichayakul, C., Kosalaraksa, P., Martinez Vargas, L., Yu, D., Watanaveeradej, V., Espinoza, F., Dietze, R., Fernando, L., Wickramasinghe, P., Duarte MoreiraJr, E., Fernando, A. D., Gunasekera, D., Luz, K., … Borkowski, A. (2022). Three-year Efficacy and Safety of Takeda's Dengue Vaccine Candidate (TAK-003). *Clinical infectious diseases : an official publication of the Infectious Diseases Society of America*, *75*(1), 107–117. https://doi.org/10.1093/cid/ciab864

[20] Patel, S. S., Winkle, P., Faccin, A., Nordio, F., LeFevre, I., & Tsoukas, C. G. (2023). An open-label, Phase 3 trial of TAK-003, a live attenuated dengue tetravalent vaccine, in healthy US adults: immunogenicity and safety when administered during the second half of a 24-month shelf-life. *Human vaccines & immunotherapeutics*, *19*(2), 2254964. https://doi.org/10.1080/21645515.2023.2254964

[21] Pintado Silva, J., Fenutria, R., Bernal-Rubio, D., Sanchez-Martin, I., Hunziker, A., Chebishev, E., Veloz, J., Kelly, G., Kim-Schulze, S., Whitehead, S., Durbin, A., Ramos, I., & Fernandez-Sesma, A. (2022). The dengue virus 4 component of NIAID's tetravalent TV003 vaccine drives its innate immune signature. *Experimental biology and medicine (Maywood, N.J.)*, *247*(24), 2201–2212. https://doi.org/10.1177/15353702231151241

[22] Durbin A. P. (2020). Historical discourse on the development of the live attenuated tetravalent dengue vaccine candidate TV003/TV005. *Current opinion in virology*, *43*, 79–87. https://doi.org/10.1016/j.coviro.2020.09.005

[23] Whitehead, S. S., Durbin, A. P., Pierce, K. K., Elwood, D., McElvany, B. D., Fraser, E. A., Carmolli, M. P., Tibery, C. M., Hynes, N. A., Jo, M., Lovchik, J. M., Larsson, C. J., Doty, E. A., Dickson, D. M., Luke, C. J., Subbarao, K., Diehl, S. A., & Kirkpatrick, B. D. (2017). In a randomized trial, the live attenuated tetravalent dengue vaccine TV003 is well-tolerated and highly immunogenic in subjects with flavivirus exposure prior to vaccination. *PLoS neglected tropical diseases*, *11*(5), e0005584. https://doi.org/10.1371/journal.pntd.0005584

[24] Sun, W., Cunningham, D., Wasserman, S. S., Perry, J., Putnak, J. R., Eckels, K. H., Vaughn, D. W., Thomas, S. J., Kanesa-Thasan, N., Innis, B. L., & Edelman, R. (2009). Phase 2 clinical trial of three formulations of tetravalent live-attenuated dengue vaccine in flavivirus-naïve adults. *Human vaccines*, *5*(1), 33–40. https://doi.org/10.4161/hv.5.1.6348

[25] Thomas, S. J., Eckels, K. H., Carletti, I., De La Barrera, R., Dessy, F., Fernandez, S., Putnak, R., Toussaint, J. F., Sun, W., Bauer, K., Gibbons, R. V., & Innis, B. L. (2013). A phase II, randomized, safety and immunogenicity study of a re-derived, live-attenuated dengue virus vaccine in healthy adults. *The American journal of tropical medicine and hygiene*, *88*(1), 73–88. https://doi.org/10.4269/ajtmh.2012.12-0361

[26] Bauer, K., Esquilin, I. O., Cornier, A. S., Thomas, S. J., Quintero Del Rio, A. I., Bertran-Pasarell, J., Morales Ramirez, J. O., Diaz, C., Carlo, S., Eckels, K. H., Tournay, E., Toussaint, J. F., De La Barrera, R., Fernandez, S., Lyons, A., Sun, W., & Innis, B. L. (2015). A Phase II, Randomized, Safety and Immunogenicity Trial of a Re-Derived, Live-Attenuated Dengue Virus Vaccine in Healthy Children and Adults Living in Puerto Rico. *The American journal of tropical medicine and hygiene*, *93*(3), 441–453. https://doi.org/10.4269/ajtmh.14-0625

[27] Watanaveeradej, V., Gibbons, R. V., Simasathien, S., Nisalak, A., Jarman, R. G., Kerdpanich, A., Tournay, E., De La Barrerra, R., Dessy, F., Toussaint, J. F., Eckels, K. H., Thomas, S. J., & Innis, B. L. (2014). Safety and immunogenicity of a rederived, live-attenuated dengue virus vaccine in healthy adults living in Thailand: a randomized trial. *The American journal of tropical medicine and hygiene*, *91*(1), 119–128. https://doi.org/10.4269/ajtmh.13-0452

[28] Simasathien, S., Thomas, S. J., Watanaveeradej, V., Nisalak, A., Barberousse, C., Innis, B. L., Sun, W., Putnak, J. R., Eckels, K. H., Hutagalung, Y., Gibbons, R. V., Zhang, C., De La Barrera, R., Jarman, R. G., Chawachalasai, W., & Mammen, M. P., Jr (2008). Safety and immunogenicity of a tetravalent live-attenuated dengue vaccine in flavivirus naive children. *The American journal of tropical medicine and hygiene*, *78*(3), 426–433.

[29] Putnak, R., Barvir, D. A., Burrous, J. M., Dubois, D. R., D'Andrea, V. M., Hoke, C. H., Sadoff, J. C., & Eckels, K. H. (1996). Development of a purified, inactivated, dengue-2 virus vaccine prototype in Vero cells: immunogenicity and protection in mice and rhesus monkeys. *The Journal of infectious diseases*, *174*(6), 1176–1184. https://doi.org/10.1093/infdis/174.6.1176

[30] Diaz, C., Lin, L., Martinez, L. J., Eckels, K. H., Campos, M., Jarman, R. G., De La Barrera, R., Lepine, E., Toussaint, J. F., Febo, I., Innis, B. L., Thomas, S. J., & Schmidt, A. C. (2018). Phase I Randomized Study of a Tetravalent Dengue Purified Inactivated Vaccine in Healthy Adults from Puerto Rico. *The American journal of tropical medicine and hygiene*, *98*(5), 1435–1443. https://doi.org/10.4269/ajtmh.17-0627

[31] Fernandez, S., Thomas, S. J., De La Barrera, R., Im-Erbsin, R., Jarman, R. G., Baras, B., Toussaint, J. F., Mossman, S., Innis, B. L., Schmidt, A., Malice, M. P., Festraets, P., Warter, L., Putnak, J. R., & Eckels, K. H. (2015). An adjuvanted, tetravalent dengue virus purified inactivated vaccine candidate induces long-lasting and protective antibody responses against dengue challenge in rhesus macaques. *The American journal of tropical medicine and hygiene*, *92*(4), 698–708. https://doi.org/10.4269/ajtmh.14-0268

[32] Danko, J. R., Kochel, T., Teneza-Mora, N., Luke, T. C., Raviprakash, K., Sun, P., Simmons, M., Moon, J. E., De La Barrera, R., Martinez, L. J., Thomas, S. J., Kenney, R. T., Smith, L., & Porter, K. R. (2018). Safety and Immunogenicity of a Tetravalent Dengue DNA Vaccine Administered with a Cationic Lipid-Based Adjuvant in a Phase 1 Clinical Trial. *The American journal of tropical medicine and hygiene*, *98*(3), 849–856. https://doi.org/10.4269/ajtmh.17-0416

[33] Danko, J. R., Kochel, T., Teneza-Mora, N., Luke, T. C., Raviprakash, K., Sun, P., Simmons, M., Moon, J. E., De La Barrera, R., Martinez, L. J., Thomas, S. J., Kenney, R. T., Smith, L., & Porter, K. R. (2018). Safety and Immunogenicity of a Tetravalent Dengue DNA Vaccine Administered with a Cationic Lipid-Based Adjuvant in a Phase 1 Clinical Trial. *The American journal of tropical medicine and hygiene*, *98*(3), 849–856. https://doi.org/10.4269/ajtmh.17-0416

[34] Manoff, S. B., Sausser, M., Falk Russell, A., Martin, J., Radley, D., Hyatt, D., Roberts, C. C., Lickliter, J., Krishnarajah, J., Bett, A., Dubey, S., Finn, T., & Coller, B. A. (2019). Immunogenicity and safety of an investigational tetravalent recombinant subunit vaccine for dengue: results of a Phase I randomized clinical trial in flavivirus-naïve adults. *Human vaccines & immunotherapeutics*, *15*(9), 2195–2204. https://doi.org/10.1080/21645515.2018.1546523

[35] Durbin, A. P., Pierce, K. K., Kirkpatrick, B. D., Grier, P., Sabundayo, B. P., He, H., Sausser, M., Russell, A. F., Martin, J., Hyatt, D., Cook, M., Sachs, J. R., Lee, A. W., Wang, L., Coller, B. A., & Whitehead, S. S. (2020). Immunogenicity and Safety of a Tetravalent Recombinant Subunit Dengue Vaccine in Adults Previously Vaccinated with a Live Attenuated Tetravalent Dengue Vaccine: Results of a Phase-I Randomized Clinical Trial. *The American journal of tropical medicine and hygiene*, *103*(2), 855–863. https://doi.org/10.4269/ajtmh.20-0042

[36] Pardi, N., Hogan, M. J., Porter, F. W., & Weissman, D. (2018). mRNA vaccines - a new era in vaccinology. *Nature reviews. Drug discovery*, *17*(4), 261–279. https://doi.org/10.1038/nrd.2017.243

[37] Roth, C., Cantaert, T., Colas, C., Prot, M., Casadémont, I., Levillayer, L., Thalmensi, J., Langlade-Demoyen, P., Gerke, C., Bahl, K., Ciaramella, G., Simon-Loriere, E., & Sakuntabhai, A. (2019). A Modified mRNA Vaccine Targeting Immunodominant NS Epitopes Protects Against Dengue Virus Infection in HLA Class I Transgenic Mice. *Frontiers in immunology*, *10*, 1424. https://doi.org/10.3389/fimmu.2019.01424

[38] Kurane, I., Zeng, L., Brinton, M. A., & Ennis, F. A. (1998). Definition of an epitope on NS3 recognized by human CD4+ cytotoxic T lymphocyte clones cross-reactive for dengue virus types 2, 3, and 4. *Virology*, *240*(2), 169–174. https://doi.org/10.1006/viro.1997.8925

[39] Livingston, P. G., Kurane, I., Dai, L. C., Okamoto, Y., Lai, C. J., Men, R., Karaki, S., Takiguchi, M., & Ennis, F. A. (1995). Dengue virus-specific, HLA-B35-restricted, human CD8+ cytotoxic T lymphocyte (CTL) clones. Recognition of NS3 amino acids 500 to 508 by CTL clones of two different serotype specificities. *Journal of immunology (Baltimore, Md. : 1950)*, *154*(3), 1287–1295.

[52] López-Sagaseta, J., Malito, E., Rappuoli, R., & Bottomley, M. J. (2015). Self-assembling protein nanoparticles in the design of vaccines. *Computational and structural biotechnology journal*, *14*, 58–68. https://doi.org/10.1016/j.csbj.2015.11.001

[53] Mohsen, M. O., & Bachmann, M. F. (2022). Virus-like particle vaccinology, from bench to bedside. *Cellular & molecular immunology*, *19*(9), 993–1011. https://doi.org/10.1038/s41423-022-00897-8

[54] Donaldson, B., Lateef, Z., Walker, G. F., Young, S. L., & Ward, V. K. (2018). Virus-like particle vaccines: immunology and formulation for clinical translation. *Expert review of vaccines*, *17*(9), 833–849. https://doi.org/10.1080/14760584.2018.1516552